\documentclass[12pt,a4paper]{JHEP3}
\input epsf
\def\vtwo{\vartheta_2}
\def\vthree{\vartheta_3}
\def\vfour{\vartheta_4}
\def\beq{\begin{equation}}
\def\eeq{\end{equation}}

\preprint{DIAS-STP-05-07}

\title{
Meromorphic Scaling Flow of $N=2$ Supersymmetric $SU(2)$ 
Yang-Mills with Matter}
\author{Brian P. Dolan\\
Dept. of Mathematical Physics, \\
National University of Ireland, Maynooth\\
and \\
Dublin Institute for Advanced Studies,\\
10, Burlington Rd., Dublin, Ireland\\
\email{\tt bdolan@thphys.nuim.ie}} 

\abstract{$\beta$-functions are derived for the flow of $N=2$ SUSY
$SU(2)$ Yang-Mills in 4-dimensions with massless matter 
multiplets in the fundamental
representation of the gauge group.
The $\beta$-functions represent the flow of the couplings as 
the VEV of the Higgs field is lowered and
are modular forms of weight -2.  They have the correct asymptotic
behaviour at both the strong and weak coupling fixed points.
Corrections to the massless $\beta$-functions when masses
are turned on are discussed.

\bigskip

}


\begin{document}

\section{Introduction}

 In a previous paper \cite{SUSYQHE} 
the flow of $N=2$ SUSY pure $SU(2)$ Yang-Mills,
with no matter fields, was analysed and a meromorphic
$\beta$-function was constructed which is finite at both weak
and strong coupling.  Up to a constant factor this $\beta$-function
reproduces the correct 1-loop Callan-Symanzik flow at both strong and 
weak coupling and interpolates between them analytically, although there 
is no a priori reason to interpret it as Callan-Symanzik $\beta$-function
away from the region of the fixed points.
This analysis modified previous suggestions
in the literature concerning the $\beta$-functions for 
$N=2$ SUSY \cite{MN, DHoker, Ritz, RDBMLL} and evades the criticisms in 
\cite{Konishi}.  

In the present paper the analysis is extended to
include massless matter fields in the fundamental representation
of $SU(2)$ with $N_f=1,2,3$ flavours.  
The construction uses the gauge invariant 
flow parameter $u=tr<\varphi^2>$, where $\varphi$ 
is the Higgs field
whose VEV is a free parameter, and the fact that the $\beta$-functions 
are modular forms of weight $-2$ of a sub-group of the full 
modular group, $\Gamma(1)\approx Sl(2,{\bf Z})/{\bf Z}_2$, 
depending on $N_f$.  The significance of the parameter $u$
was emphasised in \cite{Matone} where is was shown that $u$ is the
Legendre transform of the pre-potential.  

Following \cite{SW1,SW2} 
a convenient choice of modular parameter for $N_f>0$ 
is
\beq
\tau={\theta\over\pi}+{8\pi i\over g^2},
\label{tau}
\eeq
where $\theta$ is the usual topological parameter labelling $\theta$-vacua and $g$ is the Yang-Mills coupling constant.
In terms of $\tau$ the relevant sub-groups of $Sl(2,{\bf Z})$
for determining the $\beta$-functions  
are: $\Gamma^0(2)$ for $N_f=0$; $\Gamma(1)$ for $N_f=1$; 
$\Gamma_0(2)$ for $N_f=2$; and $\Gamma_0(4)$ for $N_f=3$.\footnote{The 
notation is that of \cite{Koblitz} and is
summarised in the appendix for ease of reference.}
For $N_f=0$ and $N_f=2$ these are larger than the monodromy group. 
The $N_f=1$ case realises the full modular group,
so the self-dual point $\tau=i$ is a fixed point of the
element sending $\tau\rightarrow -{1\over\tau}$ which is in 
the monodromy group. 
All the $\beta$-functions discussed here do still have singularities
somewhere in the fundamental domain of the relevant 
sub-group of $\Gamma(1)$,
they must do since any modular form of weight $-2$ must have
at least one singularity within, or on the boundary of, the fundamental
domain,
but these singularities are off the real axis and, with the
exception of $N_f=1$, correspond
to repulsive fixed points in both directions of the flow. 
For the $N_f=1$ $\beta$-functions 
there are two types of singularities off the
real axis: one at $\tau=i$ and its images under $\Gamma(1)$
which is repulsive in both directions of the flow;
and one at $\tau=e^{i\pi/3}$ and its images under $\Gamma(1)$
which is attractive in the direction of decreasing Higgs VEV.

The strategy is to use the technique of \cite{MN} where the
Seiberg-Witten curves describing the various theories \cite{SW2}
are written both in terms of the \lq bare' $\tau=i\infty$ and 
the renormalised finite $\tau$ and the co-efficients compared
to extract $\tau(u)$.  An important tool in the analysis
is the modular symmetry derived in \cite{SW1,SW2} where it was
shown that $N=2$ SUSY Yang-Mills has an infinite hierarchy
of vacua with massless BPS states and the modular group relates
these vacua to each other.  The value of $\tau$ in one vacuum is related
to that of another by
\beq
\tau \rightarrow \gamma(\tau)={a\tau +b \over c\tau +d}
\eeq
where $\gamma=\pmatrix{a&b\cr c&d \cr}\in\Gamma\subset \Gamma(1)$
with $\Gamma$ a sub-group of the full modular group $\Gamma(1)\cong
PSl(2,{\bf Z})$.
We therefore have, under a variation $\delta \tau$ of $\tau$,
\beq
\delta\gamma(\tau)={1\over (c\tau +d)^2}\delta\tau,
\eeq
since $ad-bc=1$, so we expect that
\beq
\beta(\gamma(\tau))={1\over (c\tau +d)^2}\beta(\tau).
\eeq
If $\beta$ is meromorphic, it will be a modular form of weight -2
and this fact proves to be a powerful analytical tool.

In \S 2 the $\beta$-function for 
$N_f=0$ discussed in \cite{SUSYQHE} is re-derived in
terms of (\ref{tau}), which differs from the normalisation in 
\cite{SUSYQHE}.  The cases $N_f=2$, $N_f=3$ and $N_f=1$ are
treated in sections \ref{S2}, \ref{threeflavours} 
and \ref{SNf1} respectively where meromorphic
$\beta$-functions are proposed which vanish at all
the strong coupling fixed points and are constant
at the weak coupling fixed point.

In principle the $\beta$-functions for different values of $N_f$ should
be related by holomorphic decoupling but the inclusion of non-zero
masses for the matter multiplets makes analytic calculations much harder
and cannot be pushed through using the techniques of the present
analysis.  In section 6 a perturbative approach is presented, 
using a strong coupling expansion and turning on a mass for one of the matter fields.  
It is shown that 
the limits $\tau_D=-1/\tau\rightarrow i\infty$
and $m\rightarrow 0$ do not commute, but nevertheless a $\beta$-function
for the massive theory with acceptable behaviour near $\tau_D=i\infty$ 
can be constructed.

 Section 7 contains our conclusions.  Two appendices give
a summary of the conventions concerning Jacobi $\vartheta$-functions
and the technical aspects of the strong coupling instanton
expansion used for the analysis in section 6.

\section{$N_f=0$}
\label{S0}

This case was treated in \cite{SUSYQHE} using the normalisation appropriate
to the adjoint representation of $SU(2)$, 
$\tilde\tau={\theta\over 2\pi}+{4\pi i\over g^2}$,
but when matter in the fundamental representation of $SU(2)$ is 
included it is better to define 
$\tau={\theta\over \pi}+{8\pi i\over g^2}$, \cite{SW2}. 
In order to set the notation and illustrate the method for $N_f\ne 0$ 
the derivation of the $\beta$-function in \cite{SUSYQHE}
is given here using the original techniques of \cite{MN},
adapted to the present  notation.

First recall the monodromy of the $N_f=0$ theory \cite{SW2}.
It is  generated by
\begin{equation}
{\cal M}_0 = \pmatrix{1 & 0 \cr -1 & 1 \cr}, \qquad
{\cal M}_{\infty} =\pmatrix{-1 & -4 \cr 0 & -1 \cr}
\end{equation}
with
\begin{equation}
{\cal M}_{\infty}{\cal M}_{0}={\cal M}^{-1}_2 \qquad
\hbox{where}\qquad
{\cal M}_2 =\pmatrix{-1 & 4 \cr -1 & 3\cr},
\end{equation}
\cite{SW2} (${\cal M}_{0}$ leaves $\tau=0$ invariant, ${\cal M}_{2}$ leaves
$\tau=2$ invariant and ${\cal M}_{\infty}$ leaves
$\tau=i\infty$ invariant).  The two matrices
${\cal M}_{\infty}$ and ${\cal M}^{-1}_{0}$,
\begin{equation}
\tau\rightarrow \tau + 4, \qquad
\tau\rightarrow {\tau\over \tau +1},
\end{equation}
generate $\Gamma^0(4)$ and therefore $\beta$ will be a modular form
for $\Gamma^0(4)$ of weight -2.
To determine its explicit form we start by following the method of \cite{MN}.  
The massless $N_f=4$ curve can be written
\begin{equation}
y^2=x^4 -2\bar u F(\tau)x^2 +\bar u^2, 
\label{Nf4}
\end{equation}
where 
\begin{equation}
F(\tau)={\vartheta_3^4(\tau)+\vartheta_2^4(\tau)\over \vartheta_4^4(\tau)}
\label{Fdef}
\end{equation}
and the $N_f=0$ curve of \cite{SW1,SW2} can be written as
\begin{equation}
y^2=x^4 -2ux^2 + u^2 -\Lambda_0^4.
\end{equation}
Note that $F(\tau)$ is an invariant function under $\Gamma(2)$.

Equation (\ref{Nf4}) is scale invariant in the sense that 
$\tau$ does not depend on 
$\bar u$.
Equating co-efficients and eliminating $\bar u$ gives 
\begin{equation}
F(\tau)={u\over \sqrt{u^2-\Lambda_0^4}}
\label{Fu}
\end{equation}
hence
\begin{equation}
u{d\tau\over du} F'(\tau)=-F(F^2-1).
\end{equation}
where $F'={dF\over d\tau}$.
Using (\ref{thetadot}) to
evaluate $F'(\tau)$ from (\ref{Fdef}) gives
\begin{equation}
F'= 2\pi i{\vartheta_3^4(\tau)\vartheta_2^4(\tau)\over\vartheta_4^4(\tau)}
\label{Fprime}
\end{equation}
finally leading to
\begin{equation}
u{d\tau\over d u}={2 i\over \pi}
{\vartheta_3^4(\tau)+\vartheta_2^4(\tau) \over \vartheta_4^8(\tau)}
\label{MNbeta}
\end{equation}
which is the result of \cite{MN}, after taking account of the
difference of notation (the variable $\tau$ here 
is $2\tau$ in \cite{MN}).
The asymptotic behaviour of (\ref{MNbeta}) is
\begin{equation}
u{d\tau\over d u}
\quad \mathop{\longrightarrow}_{\tau\rightarrow i\infty}\quad  {2i\over \pi}
\end{equation}
which is the correct asymptotically free behaviour.
However it was pointed out in \cite{Konishi} that (\ref{MNbeta}) has a
pathology in that it is singular at strong coupling,
in particular at $\tau=0$ ($u=\Lambda_0^2$) and $\tau=2$ ($u=-\Lambda_0^2$).  

A remedy was proposed
in \cite{SUSYQHE}, based on \cite{Ritz}.   The idea is to tame the singularities 
at strong
coupling, without disturbing the asymptotic properties at $u\rightarrow\infty$,
by defining
\begin{equation}
\beta(\tau)=-{(u-\Lambda_0^2)^m(u+\Lambda_0^2)^n\over u^{m+n}}u{d\tau\over d u},
\label{mnform}
\end{equation}
where $m$ and $n$ are positive integers 
(the sign is chosen so the direction of flow is the same as in
\cite{MN}).  
The correct behaviour at strong coupling then forces $m=n=1$.

The motivation behind (\ref{mnform}) relies on a theorem that
any modular form of weight -2 for a sub-group $\Gamma\subset\Gamma(1)$ can always be written as
\beq
\beta(\tau)={P(f)\over Q(f) f'}
\eeq
where $f(\tau)$ is a particular invariant function for $\Gamma$ 
(essentially one with the smallest number of zeros plus number of poles,
counting multiplicity) 
and $P(f)$ and $Q(f)$
are polynomials in $f$ (see {\it e.g} \cite{Rankin}, page 111, the idea of applying this
theorem to $N=2$ SUSY was first proposed in \cite{Ritz}).
Now $u(\tau)$ is just such an invariant function
for $\Gamma^0(4)$ --- the explicit form of $u(\tau)$ 
follows from (\ref{Fdef}) and (\ref{Fu})
\begin{equation}
{u\over \Lambda_0^2}={\vartheta_3^4(\tau) + \vartheta_2^4(\tau)\over
2\vartheta_3^2(\tau)\vartheta_2^2(\tau)},
\label{uNf2}
\end{equation}
and it can be verified explicitly, using (\ref{Tshift}) and (\ref{Sduality}),  
that $u$ is invariant under $\Gamma^0(4)$.  
Choosing the zeros of $P$ and $Q$ so as to get the correct
asymptotic behaviour at $\tau=0,2$ and $i\infty$ without 
introducing unnecessary
zeros or poles, one is led to (\ref{mnform}) and, using $m=n=1$
in conjunction with (\ref{MNbeta}) and (\ref{uNf2}), this leads uniquely to 
\begin{equation}
\beta(\tau)=-{(u-\Lambda_0^2)(u+\Lambda_0^2)\over u}{d\tau\over du}
={2\over \pi i}{1\over \vartheta_3^4(\tau) + \vartheta_2^4(\tau)}.
\label{Nf0}
\end{equation}

The asymptotic behaviour is
\begin{eqnarray}
\beta(\tau)&\rightarrow&{2\over \pi i}\quad \hbox{as}\quad 
\tau\rightarrow i\infty \nonumber \\
\beta(\tau)&\rightarrow&
-{1\over\pi i}\tau^2 \quad \hbox{as}\quad  \tau\rightarrow 0 \\
\beta(\tau)&\rightarrow&
-{1\over\pi i}(\tau-2)^2 \quad \hbox{as}\quad  \tau\rightarrow 2. 
\nonumber
\end{eqnarray}
In particular the behaviour $\beta\approx {i\over\pi}\tau^2$ as $\tau\rightarrow 0$
implies that $\beta(\tau_D)\approx {i\over\pi}$ as $\tau_D\rightarrow i\infty$, 
where $\tau_D=-1/\tau$ is the dual coupling.
Had any value of $n$ been used in (\ref{mnform}) other than unity, the result
near $\tau\approx 0$ would have been 
\beq
\beta(\tau_D)\propto  \left(e^{2i\pi\tau_D}\right)^{n-1}{i\over\pi}
\eeq
which is not the correct asymptotic behaviour.  
An exactly parallel argument applies at $\tau=2$
with $n$ replaced by $m$.

Near both the weak and the strong coupling fixed points
this $\beta$-function can be interpreted as a Callan-Symanzik $\beta$-function 
for $N_f=0$ SUSY $SU(2)$ Yang-Mills \cite{SUSYQHE} because
$u$ is a mass squared at weak coupling, where 
\beq 
\beta\approx -u{d\tau\over du},
\eeq
while at strong coupling near $\tau\approx 0$,
where
$u-\Lambda_0^2\approx \Lambda_0 a_D$ with $a_D$ proportional to
the mass of the BPS monopole, we have
\beq
\beta\approx -2(u-\Lambda_0^2){d\tau\over du}\approx -2a_D {d\tau\over da_D}.
\eeq

The flow is shown in figure 1.  
As the Higgs VEV is lowered there 
are attractive fixed points for all
even integral $\tau$ and repulsive fixed points for all odd integral 
$\tau$.  There are repulsive
fixed points in both flow directions at $\tau= k+i$ for all odd $k$
($\beta(\tau)$ diverges at these points because 
$\vartheta_3^4(\tau)=-\vartheta^4_2(\tau)$ there).
From the transformation properties of the $\vartheta$-functions
in the appendix (\ref{Tshift}) and (\ref{Sduality}), 
the $\beta$-function in (\ref{Nf0})
is a modular form of weight -2 for the group generated by
\begin{equation}
\tau\rightarrow \tau + 2, \qquad
\tau\rightarrow {\tau\over \tau +1},
\end{equation}
which is the group $\Gamma^0(2)$. 
$\Gamma^0(4)$ is a sub-group of $\Gamma^0(2)$
and the fact that the $\beta$-function has a larger symmetry
than that of the monodromy group is due to the ${\bf Z}_2$ 
symmetry of $\beta$ under
$u\rightarrow -u$, enforced by choosing $m=n$ in (\ref{mnform}).
This ${\bf Z}_2$ is the anomaly-free residue of global $R$-symmetry
\cite{SW2} which is not a symmetry of the effective action. 
Although $u$ is not invariant under $\Gamma^0(2)$, it changes sign under
$\tau\rightarrow \tau+2$, $u^2$ is invariant.

There are attractive fixed points at all the images of $\tau=0$
under $\Gamma^0(2)$ and repulsive fixed points at all the images of 
$\tau=1$ under
$\Gamma^0(2)$: at strong coupling all rational values $\tau=q/m$
with $q$ even are attractive fixed points and those with
$q$ odd are repulsive (with $q$ and $m$ mutually prime)
as the Higgs VEV is decreased.

The $\beta$-functions in \cite{SUSYQHE},
using $\tilde\tau=\tau/2$ rather than $\tau$, are the same as (\ref{Nf0}) above
as can be checked using (\ref{duplication}) from which
\begin{equation}
\beta(\tilde\tau)={1\over 2}\beta(\tau)={2\over \pi i}
{1\over \vartheta_3^4(\tilde\tau) + \vartheta_4^4(\tilde\tau)}.
\label{betatilde}
\end{equation}
These are modular forms for $\Gamma_0(2)$ which is generated by
\begin{equation}
\tilde\tau\rightarrow \tilde\tau +1, \qquad
\tilde\tau\rightarrow {\tilde\tau\over 2\tilde\tau +1},
\end{equation}
equivalent to
\begin{equation}
\tau\rightarrow \tau + 2, \qquad
\tau\rightarrow {\tau\over \tau +1}
\end{equation}
with $\tau=2\tilde\tau$.  The details of the flow generated by 
(\ref{betatilde}) were analysed in \cite{SUSYQHE}.

\section{$N_f=2$}
\label{S2}

The results of the last section can be immediately be used to guess
the form of the $\beta$-function for $N_f=2$, which can then be verified 
explicitly.  The monodromy for the $N_f=2$ case is generated by
\cite{SW2}
\begin{equation}
{\cal M}_0=\pmatrix{1&0\cr -2&1\cr}, \qquad
{\cal M}_\infty=\pmatrix{-1&-2\cr 0&-1\cr},
\end{equation}
with
\begin{equation}
{\cal M}_{\infty}{\cal M}_{0}={\cal M}^{-1}_1 \qquad
\hbox{where}\qquad
{\cal M}_1 =\pmatrix{-1 & 2 \cr -2 & 3\cr}
\label{Gamma2}
\end{equation}
(${\cal M}_0$ leaves $\tau=0$ invariant, ${\cal M}_1$ leaves
$\tau=1$ invariant  and ${\cal M}_\infty$ leaves 
$\tau=i\infty$ invariant).  The two matrices
${\cal M}_\infty$ and ${\cal M}^{-1}_0$,
\begin{equation}
\tau\rightarrow \tau + 2, \qquad
\tau\rightarrow {\tau\over 2\tau +1},
\end{equation}
generate $\Gamma(2)$. 
If a $\beta$-function is constructed which is invariant
under the anomaly-free ${\bf Z}_2$ acting on the $u$-plane,
$u\rightarrow -u$, along the same lines as before 
then we expect it will have a further symmetry under
\begin{equation}
\tau\rightarrow \tau + 1, \qquad
\end{equation}
and so will be a modular form of $\Gamma_0(2)$
of weight -2.  Demanding the correct behaviour at $\tau=i\infty$,
$\tau=0$ and $\tau=1$ leaves only one possibility and that is 
(\ref{betatilde}) with $\tilde\tau$ replaced by $\tau$, namely
\begin{equation}
\beta(\tau)={2\over \pi i}{1\over \vartheta_3^4(\tau) + \vartheta_4^4(\tau)},
\label{Nf2}
\end{equation}
the normalisation being determined by the asymptotic condition
\begin{equation}
\beta(\tau)\quad \mathop{\longrightarrow}_{\tau\rightarrow i\infty}\quad
{1\over \pi i}.
\end{equation}

We now verify (\ref{Nf2}) by explicit calculation using the
same technique as in the previous section.
The analysis initially parallels that of \cite{MN}:
the massless $N_f=2$ curve is
\begin{equation}
y^2=x^4 -2(u+3\Lambda_2^2/8)x^2 +(u-\Lambda_2^2/8)^2.
\end{equation}
Equating co-efficients with (\ref{Nf4}) gives
\begin{equation}
F(\tau)={u+3\Lambda_2^2/8\over u-\Lambda_2^2/8}
\end{equation}
or equivalently
\beq
u(\tau)={\Lambda_2^2\over 8}{\vartheta^4_3(\tau)\over \vartheta^4_2(\tau)}.
\label{uNf3}
\eeq
Using (\ref{Fdef}) and (\ref{Fprime}), now leads to
\begin{equation}
u{d\tau\over d u}=-{(F-1)(F+3)\over 4F'}={i\over 2\pi}
{\vartheta_3^4(\tau)+\vartheta_4^4(\tau)\over \vartheta_3^4(\tau) \vartheta_4^4(\tau)}
\end{equation}
as in \cite{MN}.  There are singularities at $\tau=0$ 
(where $u=\Lambda_2^2/8$) and $\tau=1$ (where $u=-\Lambda_2^2/8$) which 
can be modified, along lines
similar to \S \ref{S0}, to give a $\beta$-function
with the correct strong coupling behaviour,
\begin{equation}
\beta(\tau) = -{\left(u^2-{\Lambda_2^4\over64}\right)\over u^2} u{d\tau\over d u}
= {2\over \pi i}
{1\over \vartheta_3^4(\tau)+\vartheta_4^4(\tau)},
\end{equation}
confirming (\ref{Nf2}).  This flow is essentially the same as
that of the $N_f=1$ case treated in \S \ref{S0}, except that
$\tau$ is rescaled by a factor of $1/2$, and the flow shown in figure 2.
 
\section{$N_f=3$}
\label{threeflavours}

The monodromy for $N_f=3$ is generated by
\begin{equation}
{\cal M}_0=\pmatrix{1&0\cr -4&1\cr}, \qquad
{\cal M}_\infty=\pmatrix{-1&-1\cr 0&-1\cr}
\end{equation}
with
\begin{equation}
{\cal M}_0{\cal M}_{\infty}={\cal M}^{-1}_{-1/2} \qquad
\hbox{where}\qquad
{\cal M}_{-1/2} =\pmatrix{3 & 1 \cr -4 & -1\cr}
\label{Gamma4}
\end{equation}
(${\cal M}_0$ leaves $\tau=0$ invariant, ${\cal M}_{-1/2}$ leaves
$\tau=-1/2$ invariant  and ${\cal M}_\infty$ leaves 
$\tau=i\infty$ invariant), \cite{SW2}.  The two matrices
${\cal M}_\infty$ and ${\cal M}^{-1}_0$,
\begin{equation}
\tau\rightarrow \tau + 1, \qquad
\tau\rightarrow {\tau\over 4\tau +1},
\end{equation}
generate $\Gamma_0(4)$.
But it is not obvious what the full
symmetry of the $\beta$-functions might be
as ${\bf Z}_2$ does not play any role in the $N_f=3$ theory \cite{SW2}.
We must therefore perform the explicit calculation.
Our starting point this time
is the massless $N_f=4$ curve in the original form of
\cite{SW2}, namely
\begin{equation}
y^2=x^3- {1\over 4}g_2(\tau)\bar u^2 x - {1\over 4}g_3(\tau)\bar u^3
\label{yx3}
\end{equation}
with
\begin{equation}
g_2(\tau)={2\over 3}\Bigl(\vtwo^8(\tau)+\vthree^8(\tau)+\vfour^8(\tau)\Bigr)
\end{equation}
and
\begin{equation}
g_3(\tau)={4\over 27}\Bigl(\vthree^4(\tau)+\vfour^4(\tau)\Bigr)
\Bigl(\vtwo^4(\tau)+\vthree^4(\tau)\Bigr)
\Bigl(\vfour^4(\tau)-\vtwo^4(\tau)\Bigr),
\end{equation}
the combination 
\begin{equation}
{g_2^3(\tau)\over g_2^3(\tau) -27 g_3^2(\tau)}=
{(\vtwo^8 +\vthree^8 +\vfour^8)^3\over 54 \vtwo^8 \vthree^8\vfour^8}=J(\tau)
\end{equation}
being Klein's absolute invariant which is invariant under the action
of $\Gamma(1)$ on $\tau$.

The massless $N_f=3$ curve is
\begin{equation}
y^2=\left(x^2-{\Lambda_3^2\over 64}(x-u)\right)(x-u).
\label{yx2}
\end{equation}
Without loss of generality we can set\footnote{This is
equivalent to defining the dimensionless variable $\tilde u=64u/(\Lambda_3^2))$ and then dropping the tilde.} 
$\Lambda_3^2=64$ 
and then eliminate the quadratic term in (\ref{yx2}) by shifting 
$x\rightarrow x+(u+1)/3$
giving
\begin{equation}
y^2=x^3 -{1\over 3}(u^2 - 4u +  1)x -
{1\over 27}( 2u-1)(u^2 +8u -2).
\label{yx}
\end{equation}
There are singular points were the roots of this equation coincide,
at $u=0$, $u=1/4$ and $|u|=\infty$, corresponding to
$\tau=0$, $\tau=-1/2$ and $\tau=i\infty$ respectively.

Equating co-efficients between (\ref{yx3}) and (\ref{yx}) and eliminating
$\bar u$ yields
\begin{equation}
J(\tau)={4\over 27}{(u^2-4u+1)^3\over u^4(1-4u)}.
\label{Ju}
\end{equation}

To extract the $\beta$-function from this equation we will need an
explicit expression for $u(\tau)$.
To this end define
\beq
Y(u)={u^2\over 1-4u} \qquad\hbox{and}\qquad 
X(\tau)={\vfour^4(\tau) \over \vthree^4(\tau)}={2\over F(\tau)+1}
\eeq
in terms of which (\ref{Ju}) reads
\beq
{(1+Y)^3\over Y^2} = {(1-X + X^2)^3\over (1-X)^2 X^2}
\eeq
with the three roots
\beq
Y_1=-X(1-X),\qquad Y_2={(1-X)\over X^2}\qquad
\hbox{and} \qquad Y_3={X\over (1-X)^2}.
\eeq
Comparing with the three asymptotic forms
\begin{eqnarray}
\tau\rightarrow i\infty, &\qquad& X\rightarrow 1,\qquad |u|\rightarrow \infty, \qquad Y\rightarrow \infty,\nonumber\\
\tau\rightarrow 0, &\qquad& X\rightarrow 0,\qquad u\rightarrow 0, \qquad Y\rightarrow 0,\\
\tau\rightarrow -1/2, &\qquad& X\rightarrow 1, \qquad u\rightarrow 1/4, \qquad Y\rightarrow \infty,\nonumber
\end{eqnarray}
only $Y_3$ has the correct asymptotic behaviour at the three singular points
and so we must choose
\beq
{u^2\over 1-4u}={X\over (1-X)^2}={\vthree^4\vfour^4\over \vtwo^8},
\eeq
which is an invariant function for $\Gamma_0(2)$ \cite{Rankin}.
Solving for $u$ the asymptotic conditions pick out the unique solution
\beq
u(\tau)=-{\vthree^2\vfour^2\over (\vthree^2-\vfour^2)^2}
\label{u3}
\eeq
which is, of course, an invariant function for $\Gamma_0(4)$.

Differentiating this equation
with respect to $\tau$, and employing (\ref{thetadot}), yields
\beq
u{d\tau\over d u}
={2i\over \pi} {1\over \bigl(\vthree^2(\tau) + \vfour^2(\tau)\bigr)^2}.
\eeq
This has the correct asymptotic form as $\tau\rightarrow i\infty$,
\beq
u{d\tau\over du}\quad\mathop{\longrightarrow}_{\tau\rightarrow i\infty}\quad
{i\over 2\pi},
\eeq
and is well behaved at $\tau=0$
\beq
u{d\tau\over du}\quad\mathop{\longrightarrow}_{\tau\rightarrow 0}\quad
-{2i\over\pi}\tau^2,
\eeq
but diverges at $\tau=-1/2$.

The method used in the previous sections for $N_f=0$ and $N_f=2$ is not immediately
applicable here
since one of the fixed points is at $u=0$: 
to eliminate the singularity at $u=1/4$ 
we would have to multiply by $(u-1/4)^m/u^m$ for some positive $m$
which would introduce another singularity at $u=0$.
One strategy is to use $m=1$ to remove the singularity at $u=1/4$
and shift the other fixed point away from the origin by shifting
$u$: thus let $u'=u-\epsilon$ for some constant $\epsilon$ and define
\beq
\beta(\tau)
=-{(u'+\epsilon -1/4)(u'+\epsilon)\over u'} {d\tau\over du'}.
\eeq
This does not disturb the behaviour at $|u|=\infty$.
Clearly this is equivalent to
\beq
\beta(\tau)
=-{(u-1/4)\over (u-\epsilon)}u{d\tau\over du}
\label{utildeform}
\eeq
which preserves the good behaviour
at $\tau=i\infty$ and $\tau=0$ and gives the correct behaviour at
$\tau=-1/2$.

In terms of $\vartheta$-functions
\beq
\beta(\tau)={1\over 2\pi i}{1\over 
\vthree^2(\tau)\vfour^2(\tau)+
\epsilon\bigl(\vthree^2(\tau) -\vfour^2(\tau)\bigr)^2},
\label{Nf3}
\eeq
and has the following asymptotic forms
\begin{eqnarray}
&\beta(\tau)& \quad \mathop{\longrightarrow}_{\tau\rightarrow \infty} \quad 
{1\over 2\pi i},\nonumber \\
&\beta(\tau)& \quad \mathop{\longrightarrow}_{\tau\rightarrow 0} \quad 
-{2\over \pi i}
{\tau^2\over \epsilon}, \label{Nfthreeasymptotic}\\
&\beta(\tau)& \quad \mathop{\longrightarrow}_{\tau\rightarrow -1/2} 
\quad -{2\over \pi i}
{(\tau+1/2)^2\over (1-4\epsilon)}. \nonumber
\end{eqnarray}
If $\epsilon$ is real and $0<\epsilon<1/4$, the 
$\beta$-function is finite at both $\tau=0$ and $\tau=-1/2$
and flows in the right direction, {i.e.} in towards the fixed points
as the Higgs VEV is lowered.

In fact just such a constant shift of $u$ was found to be necessary
in instanton calculations performed to check the validity of the
Seiberg-Witten curve \cite{AHSW,DKM}.  These instanton calculations
give $\epsilon=4/27$ when $\Lambda_3/8$ is set to one as we have
done here.  The point is that, since $N_f=3$ has no discrete
symmetry acting on the $u$-plane, there is no a priori
way, using the techniques in \cite{SW2}, to determine where the
origin of the $u$-plane should lie and a constant finite shift of
$u$ does not affect the weak coupling physics.  Seiberg and Witten
chose $u=0$ to correspond to $\tau=0$ but the instanton calculations
show that $u'=-4/27$ is a more natural choice of origin.  

The flow (\ref{Nf3}) with $\epsilon=4/27$ is plotted in figure 3.
At first glance it looks very like the flow for $N_f=0$ and $N_f=2$,
with $\tau$ rescaled, but closer examination reveals subtle differences.
Figure 3 is not symmetric under $\tau\rightarrow \tau+ 1/2$, it is slightly
distorted and the repulsive fixed point close to $\tau=(1+i)/4$
is not exactly at $\tau=(1+i)/4$, it is displaced
away from the top of the semi-circular arch by a small amount.
The flow looks like a distorted version of the $N_f=2$ flow with
$\tau$ rescaled by a factor of $2$, $\tau_{N_f=2} = 2\tau_{N_f=3}$.
This is because $\Gamma^0(4)$ acting on $\tau$ is equivalent
to $\Gamma(2)$ acting on $2\tau$: the $N_f=3$ $\beta$-functions 
for $2\tau$ are therefore modular forms of $\Gamma(2)$.  
Modular forms of $\Gamma(2)$ can be obtained by distorting modular
forms of $\Gamma_0(2)$.  Provided $0<\epsilon<1/4$ the unstable
fixed point of the $N_f=3$ flow lies on the semi-circle
in the upper-half $\tau$-plane spanning the two points
$\tau=0$ and $\tau=1/2$ on the real axis.  The special case $\epsilon=1/8$
has a higher symmetry than other values because this corresponds to the
unstable fixed point being at the top of the semi-circular arch
at $\tau=(1+i)/4$ and this gives $\Gamma_0(2)$ symmetry
acting on $2\tau$.  Other values of $\epsilon$ have the lower
symmetry of $\Gamma(2)$, as in figure 3.
Flow diagrams like figure 3 have been postulated for the quantum
Hall effect when the electron spins are poorly split \cite{Crossover}.

\section{$N_f=1$}
\label{SNf1}

The $N_f=1$ case has been left till last because it
is more involved than the three cases already considered,
though paradoxically it has a higher symmetry --- the monodromy generates
the full modular group $\Gamma(1)$.  The monodromy is calculated in
\cite{SW2}: there are four singular points, at $\tau=0$, $\tau=1$, 
$\tau=2$ and $\tau=i\infty$, with monodromies
\begin{eqnarray}
{\cal M}_0&=&\pmatrix{ 1 & 0\cr -1 & 1\cr},\qquad
{\cal M}_1=\pmatrix{ 0 & 1\cr -1 & 2\cr},\nonumber\\
{\cal M}_2&=&\pmatrix{ -1 & 4\cr -1 & 3\cr},\quad 
{\cal M}_\infty=\pmatrix{ -1 & -3\cr 0 & -1\cr}
\label{Nf1monodromy}
\end{eqnarray}
respectively.
Now
\beq
{\cal M}^2_0{\cal M}_1=\pmatrix{0&1\cr -1&0\cr}
\eeq
and 
\beq
{\cal M}_0{\cal M}_1{\cal M}^{-1}_0=\pmatrix{1&1\cr 0&1\cr}
\eeq
so the two operations
\beq
\tau\rightarrow -1/\tau \qquad\hbox{and}\qquad \tau \rightarrow \tau +1
\eeq
are in the group generated by (\ref{Nf1monodromy}), which is therefore
the full modular group $\Gamma(1)$.

Again the discussion starts along the lines of \cite{MN}.
We take the massless $N_f=4$ curve
\begin{equation}
y^2=x^3- {1\over 4}g_2(\tau)\bar u^2 x - {1\over 4}g_3(\tau)\bar u^3
\label{yx3again}
\end{equation}
with
\begin{equation}
{g_2^3(\tau)\over g_2^3(\tau) -27 g_3^2(\tau)}=
{(\vtwo^8 +\vthree^8 +\vfour^8)^3\over 54 \vtwo^8 \vthree^8\vfour^8}=J(\tau),
\end{equation}
and compare this with the massless $N_f=1$ curve
\beq
y^2=x^2(x-u)-\Lambda_1^6/64.
\eeq 
First eliminate the $x^2$ term in the $N_f=1$ curve by shifting  
$x\rightarrow x+u/3$,
\beq
y^2=x^3 -{u^2\over 3}x-\left({2\over 27}u^3 +{\Lambda_1^6\over 64}\right).
\eeq 
Equating co-efficients with (\ref{yx3again}) and eliminating $\bar u$ then produces
\beq
J(\tau)=-{1\over 4}{u^6 \over u^3+1},
\label{Ju1}
\eeq
where the scale has been set by choosing $27\Lambda_1^6/256=1$.
As $\tau\rightarrow i\infty$, $J\rightarrow\infty$ and this corresponds
to $u\rightarrow -\infty$.   But $J$ is invariant under $\Gamma(1)$ and
so $J\rightarrow\infty$ at the three points $\tau=0$, $\tau=1$ and $\tau=2$
as well and these correspond to the three roots of $u^3=-1$.
Differentiating equation (\ref{Ju1}) then leads to
\beq
u{d\tau\over du}=3\left(u^3+2\over u^3+1\right) {J\over J'}
\label{Nf1MN}
\eeq
and (\ref{thetadot}) gives
\beq
{J\over J'}={1\over 2\pi i}{\vtwo^8+\vthree^8+\vfour^8\over 
(\vthree^4+\vtwo^4)(\vtwo^4-\vfour^4)(\vthree^4+\vfour^4)}.
\eeq
As $\tau\rightarrow i\infty$, 
\beq
J/J'\rightarrow {i\over 2\pi} \qquad \hbox{and} \qquad u{d\tau\over du}={3i\over 2\pi},
\eeq
which is the correct asymptotic behaviour.  But $J/J'$ has the same value
at $\tau=0,1$ and $2$ as at $\tau=i\infty$ and $u^3=-1$ at these 3 points, 
so (\ref{Nf1MN}) diverges at
strong coupling.  Following the same procedure as before the singularities
in (\ref{Nf1MN}) can be eliminated, without disturbing the behaviour at $u\approx-\infty$, by using
\beq
\left({u^3+1\over u^3}\right)u{d\tau\over du}=3\left(u^3+2\over u^3\right) {J\over J'}.
\label{Nf1Ju}
\eeq
This is not a modular form for $\Gamma(1)$ however, since $u$ is not invariant.
Solving (\ref{Ju1}) for $u$ gives
\beq
u^3=-2\left(J\pm \sqrt{J(J-1)}\;\right).
\eeq
and both roots are necessary: the upper sign for $u\rightarrow -\infty$
and the lower sign for $u^3\approx -1$.  Eliminating $u^3$ from
(\ref{Nf1Ju}) then gives an ambiguity in the direction of flow
\beq
\left({u^3+1\over u^3}\right)u{d\tau\over du}=\pm 3{\sqrt{J(J-1)}\over J'}
={\pm 3\over i\pi\sqrt{2(\vtwo^8+\vthree^8+\vfour^8)}}.
\label{Nf1pm}
\eeq

The resolution of this problem is that the true $\beta$-function,
which should be a modular form for $\Gamma(1)$, must have poles or zeros
that are not accounted for in (\ref{Nf1Ju}).  This equation was derived
by making the minimal modification of (\ref{Nf1MN}) that would remove
the infinities at $u^3=-1$ yet not disturb the asymptotic behaviour at
$u=-\infty$.   For $N_f=1,2$ and $3$ this minimalist approach worked.
For $N_f=1$ it does not, there must be another pole or zero somewhere.  
In fact (\ref{Nf1pm}) already has a singularity at $\tau=e^{i\pi/3}$,
where $\vtwo^8+\vthree^8+\vfour^8=0$ corresponding to $u=0$, 
and its images.
This is
a fixed point of $\Gamma(1)$, meaning that there is at least one
element $\gamma\in\Gamma(1)$ that leaves it invariant.
\footnote{Actually there two such elements: 
$\gamma=\pmatrix{1 & -1\cr 1 & 0 \cr}$ and 
$\gamma=\pmatrix{0 & -1\cr 1 & -1 \cr}$.} 
The point $\tau=i$, where $\vtwo^2=\vfour^2$, is also a fixed point of $\Gamma(1)$ and, if the flow
commutes with $\Gamma(1)$, then $\tau=i$ should be a fixed point of the
flow also.  The next simplest assumption that can be made is that the
only fixed points above the real axis are at $\tau=i$ and $\tau=e^{i\pi/3}$
and their images under $\Gamma(1)$, and we shall therefore assume
that these are the only places above the real axis 
where a pole or a zero of the
$\beta$-function is allowed.  We shall further assume that the total number
of poles plus zeros (counting multiplicity) is the smallest
number compatible with these requirements, since this was our experience
for $N_f=0,1$ and $3$ and was an assumption in \cite{SW1}. Jacobi's invariant 
function takes the
values $J=0$ at $\tau=e^{i\pi/3}$ and $J=1$ at $\tau=i$.
Therefore, with the above assumption, the $\beta$-function
must be of the form
\beq
\beta(\tau)=-3{J^m(J-1)^n\over J'}
\eeq
with $m$ and $n$ integers.  In order that the $\beta$-function
has the correct asymptotic form as $\tau\rightarrow i\infty$ it must
be the case that $m+n=1$ and minimising the total number of zeros
and poles leaves only two possibilities $m=0$, $n=1$ or $m=1$, $n=0$.
Examining these two possibilities the first gives 
\beq
{J-1\over J'}={2\over \pi i}{(\vthree^4+\vtwo^4)(\vtwo^4-\vfour^4)
(\vthree^4+\vfour^4)\over (\vtwo^8+\vthree^8+\vfour^8)^2}
\eeq
which is singular when $\vtwo^8+\vthree^8+\vfour^8=0$,
that is at $\tau=e^{i\pi/3}$ and its images,
and the second gives
\beq
{J\over J'}={1\over 2\pi i}{\vtwo^8+\vthree^8+\vfour^8
\over (\vthree^4+\vtwo^4)(\vtwo^4-\vfour^4)
(\vthree^4+\vfour^4)}
\eeq
which is singular when $(\vthree^4+\vtwo^4)(\vtwo^4-\vfour^4)(\vthree^4+\vfour^4)=0$, that is at
$\tau=i$ and its images.\footnote{$\vtwo^4=\vfour^4$ at $\tau=i$,
$\vthree^4=-\vfour^4$ at $\tau=(1+i)/2$ and $\vthree^4=-\vtwo^4$ at 
$\tau=1+i$.}  
The latter has a milder singularity, and clearly has a smaller number
of poles plus zeros than the former, and so is 
the unique choice that fits the criteria.
So we conjecture that the analytic $\beta$-function for $N_f=1$
is
\beq
\beta(\tau)=-\left({u^3+1\over u^3+2} \right)u{d\tau\over du}=-{3\over 2\pi i}{(\vtwo^8+\vthree^8+\vfour^8)
\over (\vthree^4+\vtwo^4)(\vtwo^4-\vfour^4)
(\vthree^4+\vfour^4)},
\eeq
which differs from (\ref{Nf1Ju}) only by using $u^3+2$ in the denominator,
rather than $u^3$.
This flow is plotted in figure 4, the pole at $\tau=i$,
corresponding to $u^3=-2$, renders this
a repulsive fixed point in both flow directions while the
zero at $\tau=e^{i\pi/3}$, corresponding to $u=0$, 
is an attractive fixed point in the direction of decreasing Higgs VEV.

\section{Massive Matter Multiplets}
\label{massivematter}

When the matter fields in the fundamental representation
of $SU(2)$ have non-zero mass the analysis is much harder.
For massive matter multiplets the Seiberg-Witten curves determine
$\tau(u,m_i,\Lambda_{N_f})$ with $i=1,\ldots,N_f$
and the $N_f-1$ case can be determined from the $N_f$ case by holomorphic
decoupling \cite{SW2}: set 
$\Lambda^{5-N_f}_{N_f-1}=m_{N_f}\Lambda^{4-N_f}_{N_f}$
and send $m_{N_f}\rightarrow\infty$ and $\Lambda_{N_f}\rightarrow 0$
keeping $\Lambda_{N_f-1}$ finite.
In principle therefore it ought to be
possible to distort figure 3 for example, by turning on one mass,
and recover figure 2 as the mass goes to infinity.
This would correspond to a family of $\beta$-functions
obtained by differentiating $\tau(u,m_i,\Lambda_{N_f})$
with respect to $u$ and considering the $m_i$ to parameterise different
$\beta$-functions: near the fixed points this corresponds to
defining $\beta$-functions
by varying the $W^\pm$-boson, and therefore also the gluino, masses 
while keeping the quark masses fixed. 

In this section we address the question of quark masses using
a strong coupling expansion, taking the $N_f=3$ case for
illustrative purposes.
The details of the analysis are
rather technical and so are relegated to appendix B,
from which we quote the relevant formulae here.

For $N_f=3$ with three different masses $m_1$, $m_2$ and
$m_3$ the Seiberg-Witten curve is uniquely determined by 
$m_1$, $m_2$, $m_3$ and  
the $\Gamma_0(4)$ invariant parameter $u$.   
With the term quadratic in $x$ eliminated the curve is
\begin{eqnarray}
y^2&=&x^3-{1\over 3}\left(u^2-4u\tilde \Lambda_3^2+\tilde \Lambda_3^4
+3(m_1^2+m_2^2+m_3^2)\tilde \Lambda_3^2-6\tilde\Lambda_3m_1m_2m_3\right)x 
\nonumber \\
&-&{1\over 27}\left\{(2u-\tilde\Lambda_3^2)
(u^2+8u\tilde\Lambda_3^2-2\tilde \Lambda_3^4
-9(m_1^2+m_2^2+m_3^2)\tilde \Lambda_3^2)\right. \hskip 3cm \\ 
&-&\left.18(u+\tilde \Lambda_3^2)\tilde \Lambda_3m_1m_2m_3 +
27\tilde\Lambda_3^2(m_1^2m_2^2+m_2^2m_3^2+m_3^2m_1^2)\right\},\nonumber
\end{eqnarray}
where, in the notation of \cite{SW2}, $\tilde\Lambda_3=\Lambda_3/8$.
In principle this
curve determines $\tau(u,m_i)$ and there are four different masses
that could be varied to define $\beta$-functions.  To connect
with the $\beta$-functions of \S \ref{threeflavours} we shall
keep $m_i$ fixed and vary only $u$.

In the massless case the strategy was to find an explicit expression
for $u(\tau)$, in terms of Jacobi $\vartheta$-functions,
and then use the properties of the $\vartheta$-functions
to determine ${du\over d\tau}$.  The problem in the massive
case is to find $u(\tau,m_i)$ and this is much harder.
We shall first simplify the problem and only consider one non-zero
mass $m_1=m$, $m_2=m_3=0$.  Even then one cannot hope for
a closed form solution.  However it is shown in appendix \ref{AppB}
that, at strong coupling where $\tau_D=-1/\tau\rightarrow i\infty$,
it is appropriate to expand in $\tilde q_D:=e^{i\pi\tau_D/2}$.
The details are somewhat technical and left to the appendix but 
for one non-zero mass there are singularities at $\tau=0$
when $u=\pm m\tilde \Lambda_3$.  Near $u=m\tilde\Lambda_3$ the expansion
\beq
u(\tau_D)=m\tilde \Lambda_3\left(1+ \alpha(m)e^{i\pi\tau_D}+\cdots\right)
\label{uexpansion}
\eeq 
is derived in appendix B
where $\alpha(m)$ is an unknown function of $m/\tilde\Lambda_3$ 
which cannot be determined
without further assumptions, but it diverges like $\sim 1/m^2$ as $m\rightarrow 0$
and $\alpha(m)\rightarrow 16$ as $m\rightarrow\infty$.  

 For $u$ infinitesimally close to $m$ the 
$\beta$-function will be of the form 
\beq
\beta(\tau)\propto -(u-m){\partial \tau\over \partial u},
\eeq
assuming the BPS monopoles have mass $\propto u-m$.
Using
\beq
{\partial u\over \partial\tau_D}\approx i\pi m \tilde\Lambda_3
\alpha(m)e^{i\pi\tau_D}\approx{i\pi}(u-m),
\eeq
gives, for $u$ near $m$,
\beq
\beta(\tau_D)\propto -(u-m){\partial\tau_D\over \partial u}
\approx {i\over \pi}, 
\eeq
which is perfectly well behaved, even in the massless limit 
$m\rightarrow 0$.
Indeed 
\beq
\beta(\tau)\propto {i\tau^2\over \pi}
\eeq
which, up to a constant, is the same behaviour as 
equation (\ref{Nfthreeasymptotic}) even though it 
is clear from (\ref{uexpansion}) that the
limits $\tau_D\rightarrow i\infty$ and $m\rightarrow 0$ do not
commute, since $\alpha(m)$ behaves as $1/m^2$ as $m\rightarrow 0$.

\section{Conclusions}

 Explicit expressions have been proposed for the
$\beta$-functions of $N=2$ SUSY $SU(2)$ Yang-Mills
with massless matter fields in the fundamental representation.
Asymptotically close to the strong and weak coupling fixed points they
co-incide with the 1-loop Callan-Symanzik $\beta$-functions, up to
a constant factor.

The $\beta$-functions are modular forms of
sub-groups of $\Gamma(1)$ for each value of $N_f$:
\begin{eqnarray}
N_f=0  && \Gamma^0(2) \cr
N_f=1  && \Gamma(1) \cr
N_f=2  && \Gamma_0(2) \cr
N_f=3  && \Gamma_0(4), 
\end{eqnarray}
for $N_f=1$ and $N_f=3$ the group is the same as the monodromy
group and for $N_f=0$ and $N_f=2$ it is larger, due to the
${\bf Z}_2$ action on the $u$-plane.

The $\beta$-functions are determined by demanding that they have the
correct asymptotic behaviour at both weak and strong coupling fixed 
points.  The relevant flows, in the direction of decreasing Higgs VEV,
are shown in figures 1,4,2 and 3 respectively.

 These functions differ from previous expressions in the literature
in that they have the correct behaviour at all the strong coupling 
fixed points.
In all cases the $\beta$-functions have a 
singularity in the interior, or on the boundary, of the
fundamental domain, corresponding to an unstable fixed point
which is repulsive in both directions of flow.  
This is a necessary consequence of their
being modular forms of weight -2.  For the $N_f=1$ case,
for which the monodromy group is the full modular group
this repulsive fixed point is at the self-dual point $\tau=i$
and its images and in this case there is also a fully attractive 
fixed point
at $\tau=e^{i\pi/3}$ and its images.  The physical significance,
if any, of this attractive fixed point is not clear.

 The case of finite masses is probably intractable using the
methods developed here, though it may well be possible to
make progress with other gauge groups by using the methods in
\cite{MN}.  Nevertheless it has been possible to show, in a strong
coupling expansion, that turning on one quark mass in the $N_f=3$ 
case still allows for a $\beta$-function with the correct
asymptotic behaviour near $\tau=0$.  Unfortunately it is
not possible to say anything about the $\beta$-function away
from $\tau=0$ because of the limitations of the technique.

It is a pleasure to thank the Perimeter Institute, Waterloo, where
this work was completed, for hospitality.  This work was supported
in part by Enterprise Ireland Basic Research Grant no. SC/2003/415.

\appendix

\section{Appendix: properties of Jacobi $\vartheta$-functions}

We collect together some useful properties of $\vartheta$-functions.
The definitions are those of \cite{WW} and most of the formulae
here are proven in that reference.  
The three Jacobi $\vartheta$-functions 
used in the text are defined as 
\begin{eqnarray}
\vartheta_2(\tau)&=&2\sum_{n=0}^\infty q^{(n+{1\over 2})^2}=
2q^{1\over 4}\prod_{n=1}^\infty\bigl(1-q^{2n}\bigr)\bigl(1+q^{2n}\bigr)^2,\\
\vartheta_3(\tau)&=&\sum_{n=-\infty}^\infty q^{n^2}=
\prod_{n=1}^\infty\bigl(1-q^{2n}\bigr)\bigl(1+q^{2n-1}\bigr)^2,\\
\vartheta_4(\tau)&=&\sum_{n=-\infty}^\infty (-1)^nq^{n^2}=
\prod_{n=1}^\infty\bigl(1-q^{2n}\bigr)\bigl(1-q^{2n-1}\bigr)^2, 
\end{eqnarray}
where $q:=e^{i\pi\tau}$.

These three $\vartheta$-functions are not independent but are related by
\begin{equation}
\vartheta_3^4(\tau)=\vartheta_2^4(\tau) + \vartheta_4^4(\tau).
\end{equation}
The following relations can be used to determine their
properties under modular transformations:
\begin{equation}
\vartheta_2(\tau + 1 )=e^{i\pi/4}\vartheta_2(\tau), \qquad
\vartheta_3(\tau+1)=\vartheta_4(\tau), \qquad
\vartheta_4(\tau+1)=\vartheta_3(\tau),
\label{Tshift}
\end{equation}
\begin{eqnarray}
\vartheta_2(-1/\tau)&=&\sqrt{-i\tau}\;\vartheta_4(\tau), \nonumber \\
\vartheta_3(-1/\tau)&=&\sqrt{-i\tau}\;\vartheta_3(\tau), \label{Sduality}\\
\vartheta_4(-1/\tau)&=&\sqrt{-i\tau}\;\vartheta_2(\tau). \nonumber
\label{Stransform}
\end{eqnarray}
The duplication formulae
\begin{eqnarray}
\vartheta_2^2(2\tau)&=& {1\over 2}\left((\vartheta_3^2(\tau)-\vartheta_4^2(\tau)\right), \nonumber \label{duplication}\\
\vartheta_3^2(2\tau)&=& {1\over 2}\Bigl(\vartheta_3^2(\tau)+\vartheta_4^2(\tau)\Bigr), \\
\vartheta_4^2(2\tau)&=& \vartheta_3(\tau)\vartheta_4(\tau),\nonumber 
\end{eqnarray}
and
\begin{equation}
\vartheta_3(4\tau)={1\over 2}\Bigl(\vartheta_3(\tau)+ \vartheta_4(\tau)\Bigr), \quad
\vartheta_2(4\tau)={1\over 2}\Bigl(\vartheta_3(\tau)- \vartheta_4(\tau)\Bigr) \quad
\end{equation}
are also useful.

At the special points $\tau=e^{i\pi/2}$ and $\tau=e^{i\pi/3}$
the $\vartheta$-functions have the values
\begin{eqnarray}
\vartheta_3^2(e^{i\pi/2})&=&\sqrt{2}\vartheta_2^2(e^{i\pi/2})=
\sqrt{2}\vartheta_4^2(e^{i\pi/2})
={2\over \pi}K\left(\sin\left({\pi\over 4}\right)\right), \\
e^{-i\pi/4}\vartheta_2^2(e^{i\pi/3})
&=&e^{-i\pi/12}\vartheta_3^2(e^{i\pi/3})=
e^{i\pi/12}\vartheta_4^2(e^{i\pi/3})={2\over\pi}
K\left(\sin\left({\pi\over 12}\right)\right), \nonumber 
\end{eqnarray}
where $K(k)$ is the complete elliptic of the second kind:
$K\bigl(\sin(\pi/4)\bigr)={1\over 4\sqrt{\pi}}(\Gamma(1/4))^2$, with
$\Gamma(1/4)\approx 3.6256$ the Euler $\Gamma$-function evaluated
at $1/4$, and $K\bigl(\sin(\pi/12)\bigr)\approx 1.5981$.

The $\vartheta$-functions have the following asymptotic forms 
\begin{eqnarray}
\tau\rightarrow i\infty&&:\quad 
\vartheta_2(\tau)\approx 2\;e^{i\pi\tau\over 4}\rightarrow 0,
\quad \vartheta_3(\tau)\rightarrow 1,\quad 
\vartheta_4(\tau)\rightarrow 1;\\
\tau\rightarrow 0&&:\quad \vartheta_2(\tau)\approx\sqrt{i\over\tau},
\quad \vartheta_3(\tau)\approx\sqrt{i\over\tau},\quad 
\vartheta_4(\tau)\approx 2\sqrt{i\over\tau}\;e^{-{i\pi\over 4\tau}}\rightarrow 0.\nonumber
\end{eqnarray}
In addition they satisfy the following
differential equations (see \cite{Rankin}, p.231, equation (7.2.17)),
\begin{eqnarray}
{\vartheta_3^\prime\over\vartheta_3}-{\vartheta_4^\prime\over\vartheta_4}
&=&{i\pi\over 4}\vartheta_2^4,\nonumber \\
{\vartheta_2^\prime\over\vartheta_2}-{\vartheta_3^\prime\over\vartheta_3}
&=&{i\pi\over 4}\vartheta_4^4,\label{thetadot} \\
{\vartheta_2^\prime\over\vartheta_2}-{\vartheta_4^\prime\over\vartheta_4}
&=&{i\pi\over 4}\vartheta_3^4.\nonumber
\end{eqnarray}

In the text $\Gamma_0(N)$ consists of matrices 
$\gamma=\pmatrix{a&b\cr c&d\cr}$ in $\Gamma(1)\approx PSl(2,{\bf Z})$
with $c\equiv 0\  \hbox{mod}\  N$, sometimes written
\beq 
\gamma\equiv\pmatrix{*&*\cr 0 &*\cr} \  \hbox{mod}\  N,
\eeq
$\Gamma^0(N)$ consists of matrices 
with $b\equiv 0 \ \hbox{mod}\  N$,
\beq 
\gamma\equiv\pmatrix{*&0\cr * &*\cr}\  \hbox{mod}\  N
\eeq
and $\Gamma(N)$ consists of matrices 
with 
\beq 
\gamma\equiv\pmatrix{1&0\cr 0 &1\cr} \  \hbox{mod}\  N.
\eeq

\section{Appendix: Massive Matter Fields for \hbox{$N_f=3$}}
\label{AppB}

The starting point is the
curve for four massive matter multiplets
in the fundamental representation \cite{SW2},
\begin{eqnarray}
y^2&=&(x^2-c_2^2 \bar u^2)(x-c_1\bar u) - 
c_2^2(x-c_1\bar u)^2\bar A -c_2^2(c_1^2-c_2^2)(x-c_1\bar u)
\bar B\nonumber \\
&+&2c_2(c_1^2-c_2^2)(c_1x-c_2^2\bar u)\bar C -c_2^2(c_1^2-c_2^2)^2\bar D,
\end{eqnarray}  
with
\beq
c_1(\tau)={1\over 2}\bigl(\vthree^4(\tau)+\vfour^4(\tau)\bigr),\qquad
c_2(\tau)={1\over 2}\bigl(\vthree^4(\tau)-\vfour^4(\tau)\bigr),
\eeq
and
\beq
\bar A=\sum_i^4\bar m_i^2,\qquad
\bar B=\sum_{i<j}\bar m_i^2\bar m_j^2,\qquad
\bar C=\bar m_1\bar m_2 \bar m_3 \bar m_4 \qquad
\bar D=\sum_{i<j<k}\bar m_i^2 \bar m_j^2 \bar m_k^2.
\eeq

First eliminate the quadratic term in $x$ by shifting 
$x\rightarrow x+(c_1\bar u+c_2^2\bar A)/3$, giving
\beq
y^2=x^3 -{1\over 3}P(\tau,\bar u,\bar m_i)x -{1\over 27}Q(\tau,\bar u,\bar m_i),
\label{Nffour}
\eeq
where
\beq
P(\tau,\bar u,\bar m_i)=
(c_1^2+3c_2^2)\bar u^2 -4c_1c_2^2\bar A\bar u +
3c_2(c_1^2-c_2^2)(c_2\bar B-2c_1\bar C)+c_2^4\bar A^2,
\eeq
is quadratic in $\bar u$ and
\begin{eqnarray}
Q(\tau,\bar u,m_i)&=&
2c_1(c_1^2-9c_2^2)\bar u^3 +3c_2^2(5c_1^2+3c_2^2)\bar A\bar u^2 
\label{Qfunc}\\
&-&2\left(6c_1c_2^4\bar A^2+9c_1c_2^2(c_1^2-c_2^2)\bar B +9c_2(c_1^2-c_2^2)(c_1^2-3c_2^2)\bar C\right)\bar u \nonumber \\
&+&2c_2^6\bar A^3 +27c_2^2(c_1^2-c_2^2)^2\bar D
+9c_2^3(c_1^2-c_2^2)(c_2\bar B-2c_1\bar C)\bar A \nonumber
\end{eqnarray}
is cubic in $\bar u$.

Next this curve can be reduced to the $N_f=3$ curve
using the holomorphic decoupling
of \cite{SW2}: send $\bar m_4\rightarrow \infty$ and 
take the \lq bare' coupling $\tau\rightarrow i\infty$,
so $c_2\rightarrow 0$ and $c_1\rightarrow 1$, keeping $c_2\bar m_4=\Lambda_3/8:=\tilde \Lambda_3$ finite.  Taking this limit, and dropping the bars on $u$, $m_1$,
$m_2$ and $m_3$, gives the massive $N_f=3$ curve,
\begin{eqnarray}
y^2&=&x^3-{1\over 3}(u^2-4u\tilde \Lambda_3^2+\tilde \Lambda_3^4
+3B\tilde \Lambda_3^2-6C\tilde\Lambda_3)x \label{SWNf3curve}\\
&-&{1\over 27}(2u-\tilde\Lambda_3^2)
(u^2+8u\tilde\Lambda_3^2-2\tilde \Lambda_3^4-9B\tilde \Lambda_3^2)
+{2\over 3}(u+\tilde \Lambda_3^2)C\tilde \Lambda_3 -D,\nonumber
\end{eqnarray}
with 
\beq
B=m_1^2+m_2^2+m_3^2,\qquad C=m_1m_2m_3, \qquad
D=m_1^2m_2^2+m_2^2m_3^2+m_3^2m_1^2.
\label{ABC}
\eeq
This reduces to equation (\ref{yx}) when the three masses are set to zero
and $\tilde \Lambda_3=1$.  Following \cite{MN} this curve is now compared
to (\ref{Nffour}) with $\bar m_4=0$, $\bar u$ independent of $\tau$,
that is equation (\ref{Nffour})-(\ref{Qfunc}) with $\bar C=0$
and
\beq
\bar A=\bar m_1^2+\bar m_2^2+\bar m_3^2,\qquad
\bar B=\bar m_1^2\bar m_2^2+\bar m_2^2\bar m_3^2+\bar m_3^2\bar m_1^2,\qquad
\bar D=\bar m_1^2 \bar m_2^2 \bar m_3^2.
\eeq
Equating co-efficients leads to
\beq
(c_1^2+3c_2^2)\bar u^2 -4c_1c_2^2\bar A\bar u +
3c_2^2(c_1^2-c_2^2)\bar B+c_2^4\bar A^2=
u^2-4u\tilde \Lambda_3^2+\tilde \Lambda_3^4
+3B\tilde \Lambda_3^2-6C\tilde\Lambda_3 \label{x1coeff}
\eeq
\begin{eqnarray}
2c_1(c_1^2-9c_2^2)\bar u^3 +3c_2^2(5c_1^2+3c_2^2)\bar A\bar u^2 
-2\left(6c_1c_2^4\bar A^2+9c_1c_2^2(c_1^2-c_2^2)\bar B\right)\bar u \nonumber \\
+2c_2^6\bar A^3 +27c_2^2(c_1^2-c_2^2)^2\bar D
+9c_2^4(c_1^2-c_2^2)\bar B\bar A \hskip 4cm \label{x0coeff}\\
=(2u-\tilde\Lambda_3^2)
(u^2+8u\tilde\Lambda_3^2-2\tilde \Lambda_3^4-9B\tilde \Lambda_3^2) -
18(u+\tilde \Lambda_3^2)C\tilde \Lambda_3 +27D. \nonumber
\end{eqnarray} 
In the massless case we can eliminate $\bar u$ from these
two equations and determine $u(\tau)$, but now we want to determine
$u(\tau,m_i)$ with $\bar u$, $\bar A$, $\bar B$ and $\bar D$
all unknown functions so there is not enough information to solve
the problem completely.  Nevertheless we can still get information
from these equations.  The symmetry group for $N_f=3$ is $\Gamma_0(4)$
and $u$, $m_i$ and $\tilde \Lambda_3$ 
should be invariants of $\Gamma_0(4)$ so the
right hand sides of (\ref{x1coeff}) and (\ref{x0coeff}) are also
invariants.   Using the
transformation properties of the $\vartheta$-functions, (\ref{Tshift})
and (\ref{Stransform}), this implies that $\bar u$, $\bar A$, 
$\bar B$, and $\bar D$ are modular forms of weights -2, -4, -8 and -12
respectively, so $\bar m_i$ have weight -2.

For simplicity we shall focus on the case of a single mass, $m_1=m$,
$m_2=m_3=0$ so 
\beq
\bar A=\bar m^2,\qquad \bar B=0,\qquad \bar D=0
\eeq
\beq
B=m^2,\qquad D=0,\qquad C=0
\eeq
and the equations simplify to
\beq
(c_1^2+3c_2^2)\bar u^2 -4c_1c_2^2\bar m^2\bar u +c_2^4\bar m^4=
u^2-4u\tilde \Lambda_3^2+\tilde \Lambda_3^4
+3m^2\tilde \Lambda_3^2\label{x1coeffm}
\eeq
\begin{eqnarray}
2c_1(c_1^2-9c_2^2)\bar u^3 +3c_2^2(5c_1^2+3c_2^2)\bar m^2\bar u^2 
-12c_1c_2^4\bar m^4\bar u
+2c_2^6\bar m^6 \hskip -8cm \nonumber \\
&=&(2u-\tilde\Lambda_3^2)
(u^2+8u\tilde\Lambda_3^2-2\tilde \Lambda_3^4-9m^2\tilde \Lambda_3^2). 
\label{x0coeffm}
\end{eqnarray} 
We now want to eliminate $\bar u$ and $\bar m$ to get $u(\tau,m)$
but there is still not enough information.  However, knowing that
$\bar u$ and $\bar m^2$ are modular forms of weight -2 and -4
respectively, we can
say something about their functional form at strong coupling.
To see how this works let us first look at the case $\bar m=0$,
where the explicit solution is given in \S \ref{threeflavours}.
Using the details there one finds
\beq
\bar u(\tau)=-{\tilde \Lambda_3^2\over \bigl(\vthree^2(\tau)-\vfour^2(\tau)\bigr)^2}
\eeq
which is indeed a modular form for $\Gamma_0(4)$ of weight -2 and
it vanishes at strong coupling, $\tau\rightarrow 0$.  Now $\tau\rightarrow
-1/\tau=\tau_D$ is not in $\Gamma_0(4)$, so $\bar u$ is not a modular
form under this transformation, rather
\beq
\bar u(\tau_D)={1\over \tau_D^2}{\tilde\Lambda_3^2\over 
\bigl(\vthree^2(\tau_D)-\vtwo^2(\tau_D)\bigr)^2}.
\eeq
Writing $\tilde q_D=e^{i\pi\tau_D/2}$ we have the strong coupling 
expansions
\beq
\bar u(\tau_D)={\tilde\Lambda_3^2\over \tau_D^2}\left(1+8\tilde q_D+40\tilde q_D^2+160\tilde q_D^3
+\cdots  \right).
\eeq
and, from (\ref{u3}),
\beq
u(\tau_D)=-{\vartheta_3^2(\tau_D)\vartheta_2^2(\tau_D)\over
\bigl(\vartheta_3^2(\tau_D)-\vartheta_2^2(\tau_D)\bigr)^2}=-4\tilde\Lambda_3^2\tilde q_D\left(1+8\tilde q_D+44\tilde q_D^2+192\tilde q_D^3
+\cdots  \right).
\label{masslessuexpansion}
\eeq
For non-zero $m$ it is consistent with all we know to assume
a similar form
\beq
\bar u(\tau_D)={\tilde\Lambda_3^2\over \tau_D^2}\left(\bar u_0+\bar u_1\tilde q_D+\bar u_2\tilde q_D^2+\bar u_3\tilde q_D^3
+\cdots  \right).
\eeq
and similarly for $\bar m^2$
\beq
\bar m^2(\tau_D)={\tilde \Lambda_3^2\over \tau_D^4}
\left(\bar a_0+\bar a_1\tilde q_D+\bar a_2\tilde q_D^2+\bar a_3\tilde q_D^3
+\cdots  \right),
\eeq
where $\bar u_k$ and $\bar a_k$ are functions of $m/\tilde\Lambda_3$,
with $\bar a_k$ vanishing for $m=0$.
A similar strong coupling expansion for $u$ has no prefactor
of $1/\tau_D^2$ because $u$ has weight zero not -2,
\beq
{u\over \tilde\Lambda_3^2}=u_0+u_1\tilde q_D+u_2\tilde q_D^2+u_3\tilde q_D^3
+\cdots .
\eeq
Using these expansions in (\ref{x1coeffm}) and (\ref{x0coeffm}),
together with
\begin{eqnarray}
c_1(\tau)&=&c_1(-1/\tau_D)=
-{\tau_D^2\over 2}\Bigl(\vthree(\tau_D)^4+\vtwo(\tau_D)^4\Bigr),\nonumber \\
c_2(\tau)&=&c_2(-1/\tau_D)=-{\tau_D^2\over 2}\vfour(\tau_D)^4,
\end{eqnarray}
we can equate powers of $\tilde q_D$ to obtain recurrence relations
between the $u_k$ and the $\bar a_k$.  Without making further assumptions
there is not enough information to determine $u(\tau,m)$ but we
can still extract useful information about the strong coupling
$\beta$-function.  At zeroth
order in $\tilde q_D$ (\ref{x1coeffm}) and (\ref{x0coeffm})
(with $\tilde \Lambda_3=1$ for simplicity) yield three possibilities:
\beq
u_0=\pm m, \bar u_0=1\mp 2m-{\bar a_0\over 4}; \qquad \hbox{and} 
\qquad u_0 = m^2 +{1\over 4}, \bar u_0=m^2-{\bar a_0+1\over 4}.
\eeq
Only the first two are relevant for
the strong coupling fixed point, $\tau=0$ at $u=0$, 
of the massless $N_f=3$ theory (in the massless case $u=1/4$ is associated with
$\tau=-1/2$ where $\tau_D=2$).
Choosing the root $u=m$, at order $\tilde q_D$ we find
a pair of equations linear in $u_1$ and $\bar u_1$ which
are degenerate.  Solving for $\bar u_1$ gives 
\beq
\bar u_1 := -{(m-2)u_1\over (2m-1)}-{\bar a_1\over 4}.
\eeq
At order $\tilde q^2_D$ the pair of linear equations for
for $u_2$ and $\bar u_2$ are parallel in the $u_2 - \bar u_2$
plane and have no solution unless they co-incide, which
only happens if
\beq
{u_1  m \over 2 m - 1}=0.
\label{u1vanishes}
\eeq
For $m=0$ this is automatic, but for $m\ne 0$ it forces
$u_1=0$.  Setting $u_1=0$ then gives one linear equation relating
$\bar u_2$ to $u_2$,
\beq
\bar u_2=-{2(m-2)u_2 + \bar a_0 (3\bar a_0 + 8m-4)\over 2(2m-1)}-{\bar a_2\over 4}
\eeq
At order $\tilde q_D^3$ the two linear equations for $u_3$ and
$\bar u_3$ are again degenerate giving only one constraint
which can be used to solve for $\bar u_3$
\beq
\bar u_3= -{(m-2)u_3+\bar a_1(3\bar a_0+4m-2)\over (2m-1)}-{\bar a_3\over 4}.
\eeq
At order $\tilde q_D^4$ one again obtains two parallel lines in the
$u_4 - \bar u_4$ plane which do not intersect unless
\beq
u_2=\pm{(\bar a_0 +8m -4)^2\over 2m}
\eeq
in which case $\bar u_4$ can be obtained as a function of $u_4$, $\bar a_0$,
$\bar a_2$ and $m$.  

One can continue but for the present purposes we have gone as far as necessary.
We are only really 
interested in $u$ and we have
\beq
u(\tau_D)=m\pm {(\bar a_0 +8m-4)^2\over 2m}\tilde q_D^2+\cdots
\eeq 
with $\bar a_0$ and undetermined function of $m$, but independent of $\tau_D$.
To get the dimensions correct we should re-instate $\tilde\Lambda_3$
and write
\beq
u(\tau_D)=m\tilde\Lambda_3\left(1+\alpha(m) e^{i\pi\tau_D}+\cdots\right)
\label{uexpansionA}
\eeq 
where
\beq
\alpha(m):=\pm {(\bar a_0\tilde \Lambda_3^2 +8m\tilde\Lambda_3-4\tilde\Lambda_3^2)^2\over 2m^2\tilde\Lambda_3^2}.
\eeq

Note that, in order to pin down the co-efficient
$u_2$ one has to go to order $\tilde q_D^4$ and examine $u_4$.
At every value of $k$ in the expansion
one gets a pair of linear equations in $u_k$ and $\bar u_k$ in terms
of $m$ and the $\bar a_{k'}$ with $k'\le k$.  For $k=1$ these equations
are degenerate and $u_1$ is not determined; for $k=2$ the equations
have no solution unless $u_1=0$ in which case they
are again degenerate and $u_2$ is undetermined; for $k=3$ the equations
are again degenerate and $u_3$  is undetermined and for $k=4$ the equations
have no solution unless $u_2$ has one of the two possible values
 shown in (\ref{uexpansionA}). 

The explicit from of $\alpha(m)$ is not needed
in the analysis, but we can fix its asymptotic form as 
$m\rightarrow\infty$ using holomorphic decoupling \cite{SW2}.  For $N_f=2$
equation (\ref{uNf3}) gives
\beq
u(\tau_D)={\Lambda_2^2\over 8}{\vartheta_3^4(\tau_D)\over\vartheta_4^4(\tau_D)}
\approx {\Lambda_2^2\over 8}\left(1+16e^{i\pi\tau_D}+\cdots\right),
\eeq
and this should agree with (\ref{uexpansionA})
as $m\rightarrow\infty$, $\Lambda_3\rightarrow 0$ with
$\Lambda_2^2=m\Lambda_3 =8m\tilde\Lambda_3$ fixed.
Hence $\alpha(m)\rightarrow 16$ as $m\rightarrow\infty$.

The other singularity of the $N_f=2$ theory, at $u=-{\Lambda_2^2\over 8}$,
is obtained by holomorphic decoupling in the weak coupling limit of the
$N_f=3$ theory by expanding around $u=-m$.

Notice that the $m=0$ expansion for $u$ in equation 
(\ref{masslessuexpansion}) contains a term linear in $\tilde q_D$
while (\ref{uexpansionA}) does not.  This is because, when $m\ne 0$,
(\ref{u1vanishes}) forces us to set $u_1=0$.
Since $\bar a_0\rightarrow 0$ as $m\rightarrow 0$,
$\alpha(m)$ diverges like $8\tilde\Lambda^2_3/m^2$ as $m\rightarrow 0$ and 
the limits $\tau_D\rightarrow i\infty$ and $m\rightarrow 0$ do not
commute.
 
 We can perform a similar expansion around $\tau=-1/2$, using $u_0=m^2+1/4$,
where $\tau_D=2+i\varepsilon$ with $\varepsilon$ small.  
The analysis is simpler than the $u_0=\pm m$ case
in that at each order, at least up to order 5
in $\tilde q_D$ which is as far as we have gone,
one simply finds a pair of linear equations in $u_k$ and $\bar u_k$ 
which can be solved in terms of $\bar a_{k'}$ with $k'<k$.
The details are omitted but one finds
\beq
u_1=u_2=u_3=0,\qquad u_4=-4{(\bar a_0-4m^2+1)^4\over (4m^2-1)^2},
\qquad u_5=-16{\bar a_1(\bar a_0-4m^2+1)^3\over (4m^2-1)^2},
\eeq
and so, for $\tau_D=2+i\varepsilon$ with $\varepsilon$ small,
\beq
u(\tau_D)=m^2+{1\over 4}-4{(\bar a_0-4m^2+1)^4\over (4m^2-1)^2}e^{-2\pi\varepsilon}
+16{\bar a_1(\bar a_0-4m^2+1)^3\over (4m^2-1)^2}e^{-5\pi\varepsilon/2}+\cdots.
\eeq
Re-instating $\tilde \Lambda_3$ now gives
\beq
u(\tau_D)=m^2+
{\tilde\Lambda_3^2\over 4}+
\tilde\Lambda_3^2\tilde\alpha(m) e^{-2\pi\varepsilon}
\cdots.
\label{uexpansionB}
\eeq
where
\beq
\tilde\alpha(m)=-
4{(\bar a_0\tilde\Lambda_3^2-4m^2+\tilde\Lambda_3^2)^4
\over (4m^2-\tilde\Lambda_3^2)^2\tilde\Lambda_3^4}.
\eeq
As $m\rightarrow \infty$ this point in the $u$-plane goes out to infinity
in the $N_f=2$ theory.
 
Equations (\ref{uexpansionA}) and (\ref{uexpansionB}) 
were the aim of this appendix and are used
in \S \ref{massivematter} in the discussion of the $\beta$-functions
for the massive $N_f=3$ theory at strong coupling.

\newpage

\vtop{\hskip -33pt
\hbox{\epsfxsize=16cm
\epsffile{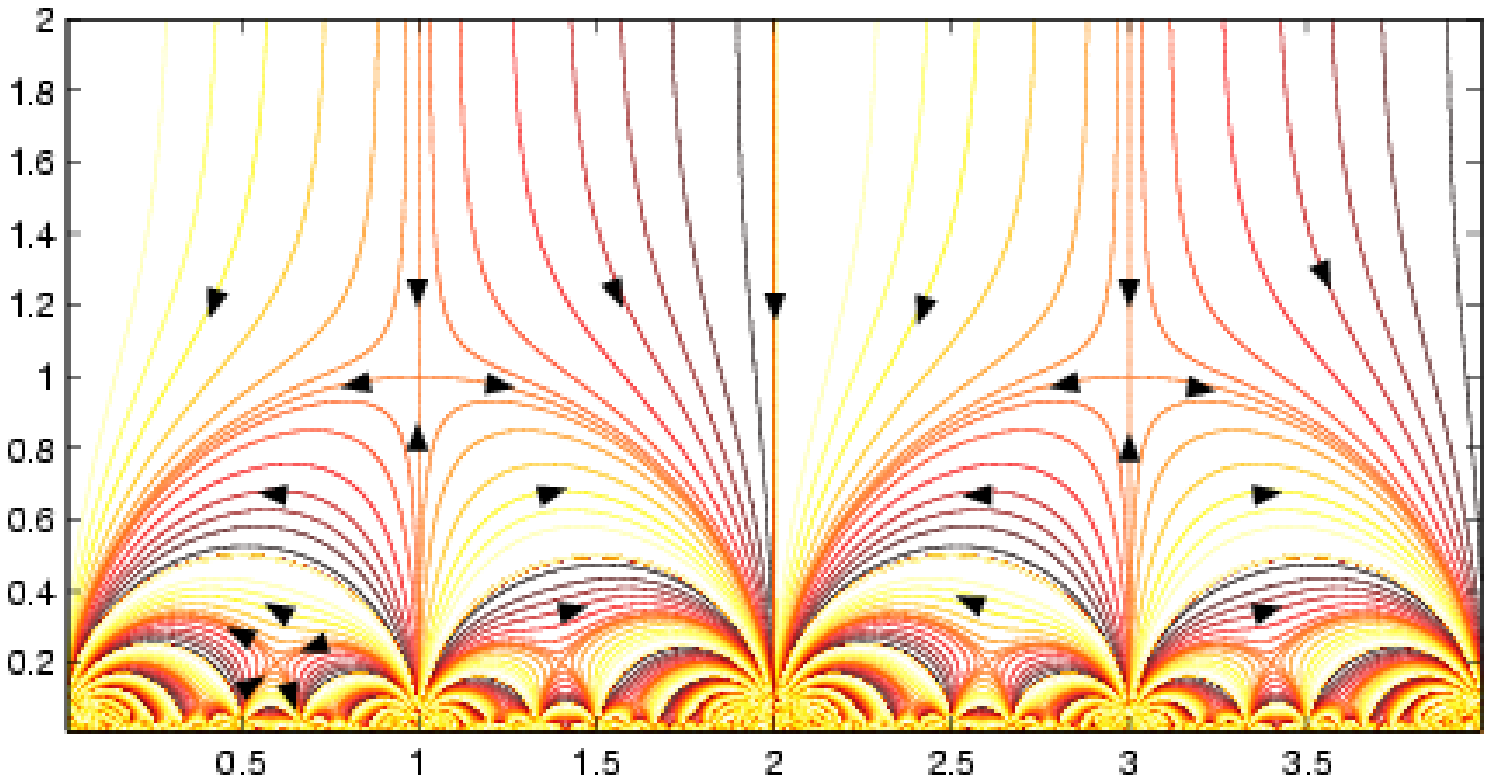}}

\noindent Fig. 1: Flow of effective coupling  
of $N=2$ SUSY Yang-Mills with $N_f=0$.  The arrows indicate
the direction of the flow as as the Higgs VEV is reduced.
The $\beta$-functions are modular forms of $\Gamma^0(2)$ and the
pattern repeats under $\tau\rightarrow \tau+2$.
}


\bigskip

\vtop{
\hbox{\epsfxsize=15cm
\epsffile{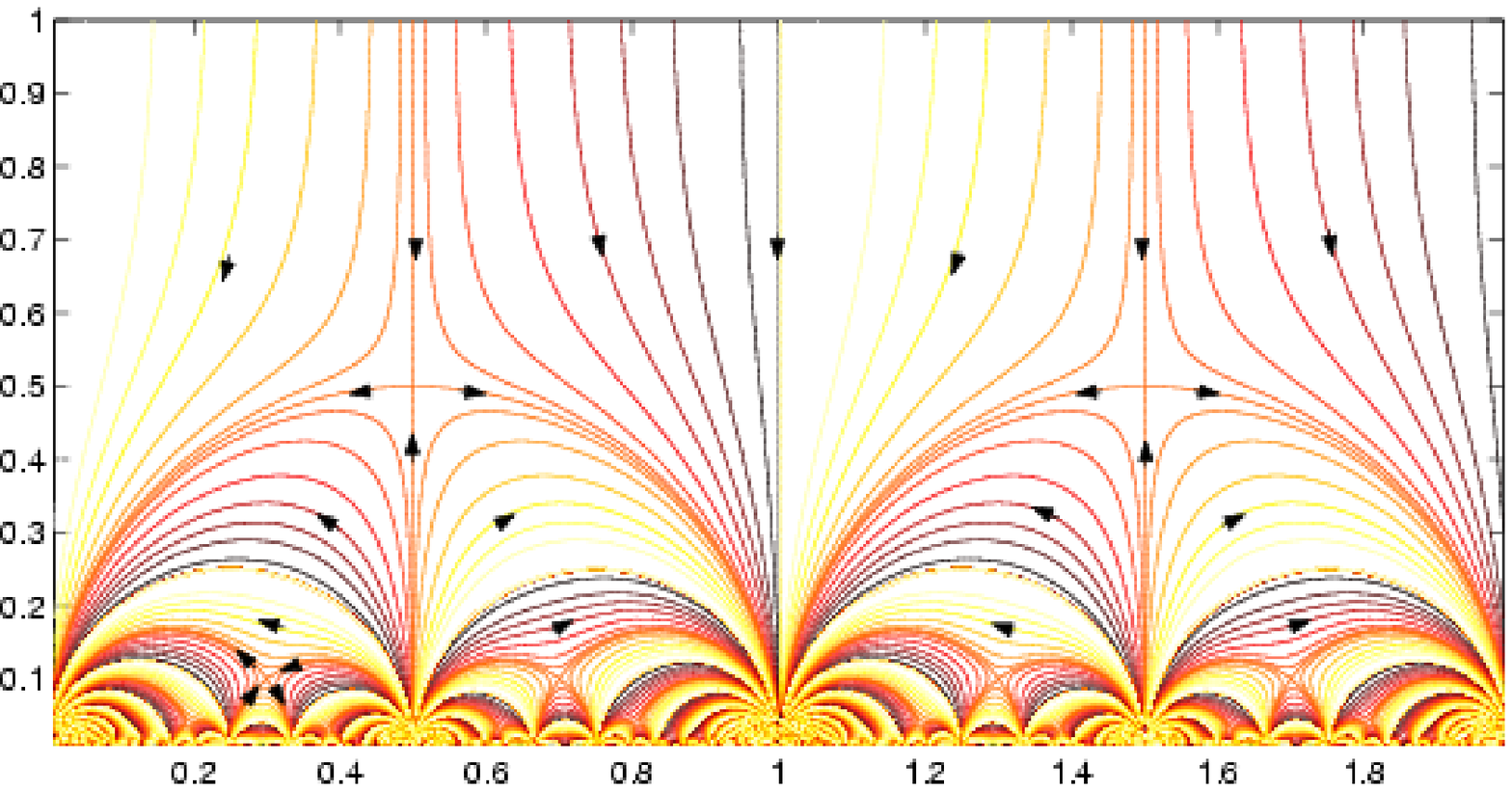}}

\noindent Fig. 2: Flow of effective coupling  
of $N=2$ SUSY Yang-Mills with $N_f=2$.  The arrows indicate
the direction of the flow as as the Higgs VEV is reduced.
The $\beta$-functions are modular forms of $\Gamma_0(2)$ and the
pattern repeats under $\tau\rightarrow \tau+1$.
}

\newpage

\vtop{
\hbox{\epsfxsize=15cm 
\epsffile{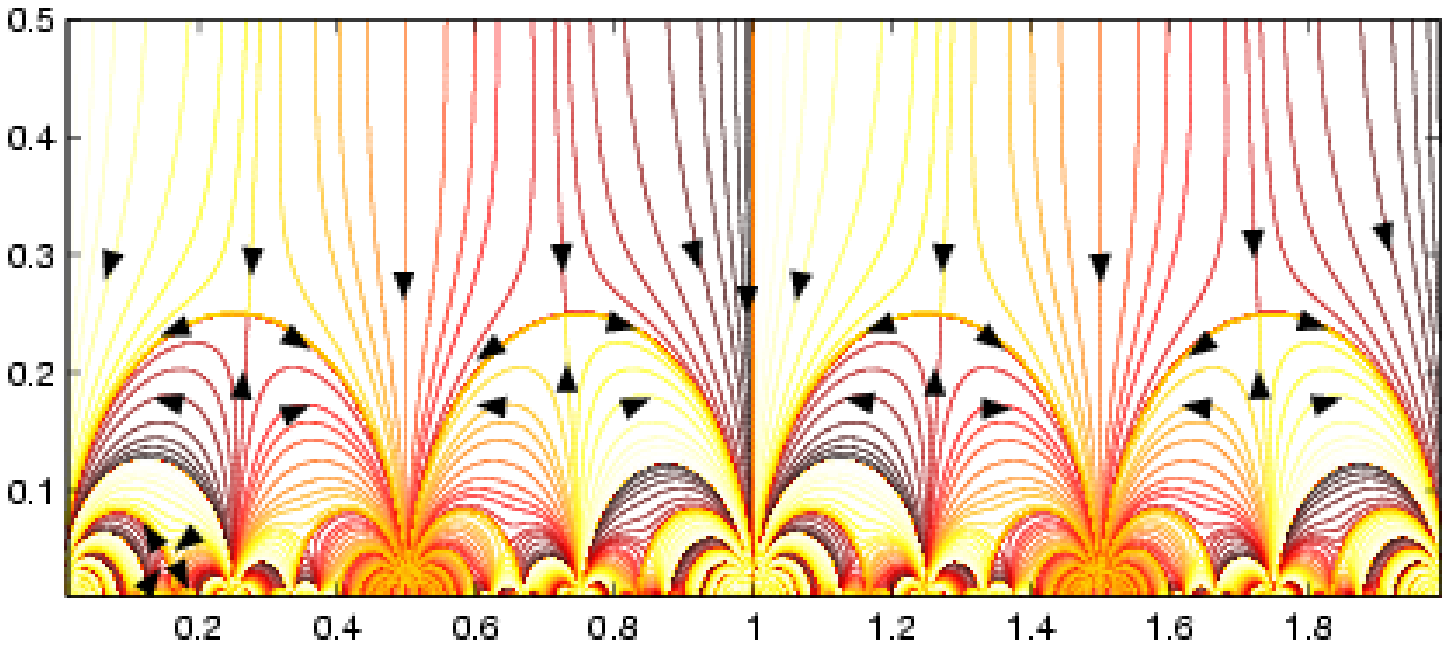}}

\noindent Fig. 3: Flow of effective coupling  
of $N=2$ SUSY Yang-Mills with $N_f=3$ and $\epsilon=4/27$.  
The arrows indicate
the direction of the flow as as the Higgs VEV is reduced.
The $\beta$-functions are modular forms of $\Gamma_0(4)$ and the
pattern repeats under $\tau\rightarrow \tau+1$.}

\vskip 1cm

\vtop{
\hskip 2cm\hbox{\epsfxsize=12cm 
\hskip -25pt \epsffile{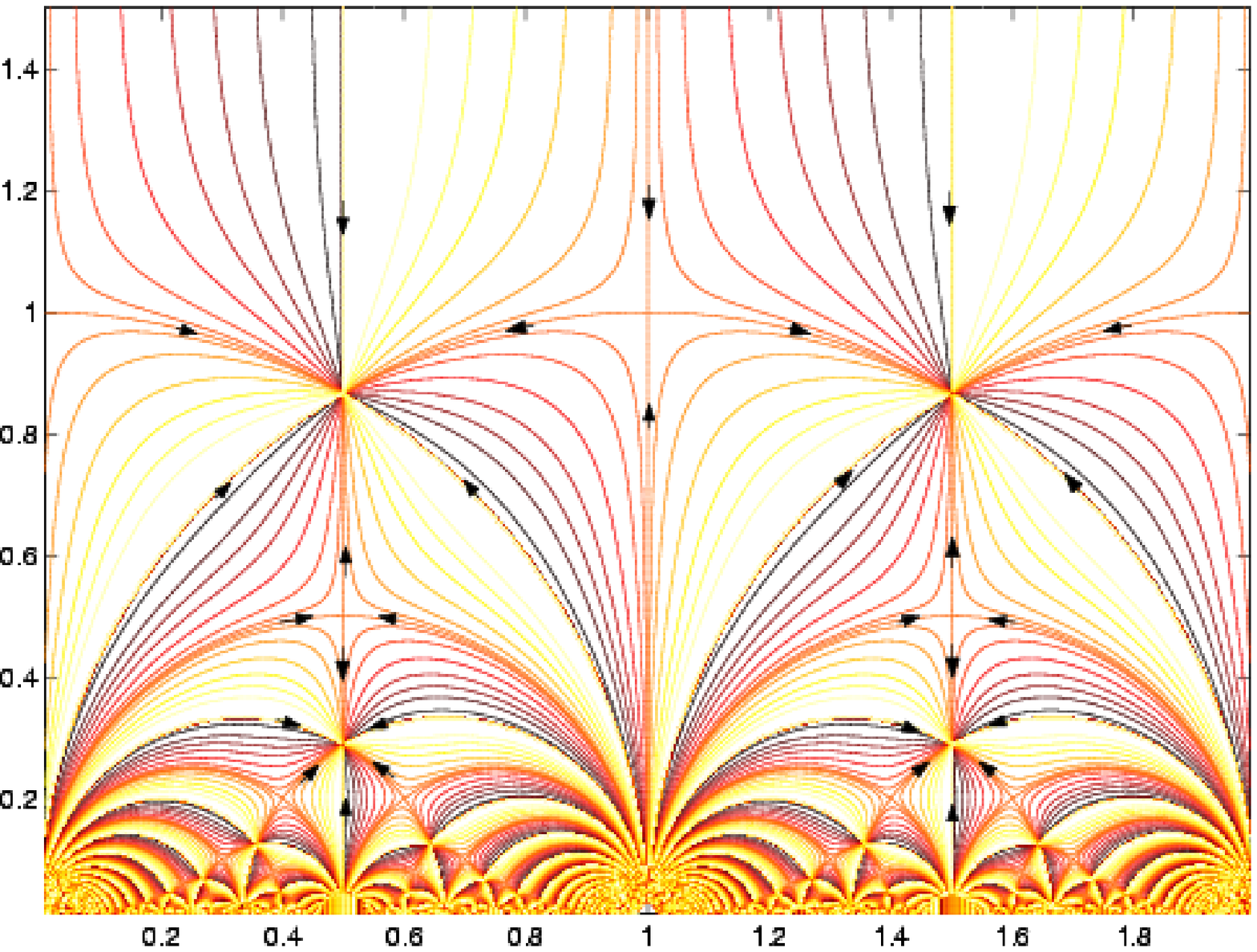}}

\noindent Fig. 4: Flow of effective coupling  
of $N=2$ SUSY Yang-Mills with $N_f=1$.  The arrows indicate
the direction of the flow as as the Higgs VEV is reduced.
The $\beta$-functions are modular forms of $\Gamma(1)$ and the
pattern repeats under $\tau\rightarrow \tau+1$.}

\end{document}